# Personal Privacy Protection Problems in the Digital Age

Zhiheng Yi[1], Xiaoli Chen[2]

Corresponding author: Zhiheng Yi, zhiheng.yi@stu.csust.edu.cn

**ABSTRACT**

With the development of Internet technology, the issue of privacy leakage has attracted more and more attention from the public. In our daily life, mobile phone applications and identity documents that we use may bring the risk of privacy leakage, which had increasingly aroused public concern. The path of privacy protection in the digital age remains to be explored. To explore the source of this risk and how it can be reduced, we conducted this study by using personal experience, collecting data and applying the theory.

## INTRODUCTION

The booming development of the Internet has not only made people have more and more novel and convenient technological experience, but also brought about problems that can not be underestimated, which has aroused wide public attention. One of the most representative problems is the leakage of personal privacy. With the Internet of things, cloud computing and big data, citizens' privacy leakage is becoming more and more rampant, making it difficult for citizens' privacy rights to be fully guaranteed. From a surveillance system at home, to a mobile phone in your hand, to a registered app account, they all can become media for personal privacy leakage. Sometimes, you even need a single phone number or a social media account to find a person's other information. In the era of big data, we are running as though we are naked. In the future, the more closely connected Internet will add to this concern. What is the future of personal privacy protection in the digital age?

---

[1] Zhiheng Yi, a Grade-1 undergraduate student in International College for Engineering, Changsha University of Science and Technology, China.
[2] Xiaoli Chen, a Grade-1 undergraduate student in Orient Science and Technology College of Hunan Agricultural University, China.

For above reason, we conducted this research. This essay discusses and studies several aspects of the privacy leakage risk in the current digital age by using personal experience, collecting data and applying the theory, and puts forward the privacy protection methods for the country, enterprises and individuals. It is hoped that these possible approaches will help personal privacy protection issues in the digital age in the future.

**DISCUSSION**

**Apps and Personal Privacy**

In 2018, the incident that Facebook provided 50 million private users to Cambridge Analytica was revealed, causing strong public opinion. As a data analytical company, the company was known for serving the Trump campaign. In 2016, it was revealed that the company acquired user information through Facebook to influence the outcome of the US election. In this case, Facebook suffered a crisis of confidence. Moreover, in recent years, Facebook's user information has been frequently leaked, making its user credit values lower than expected.

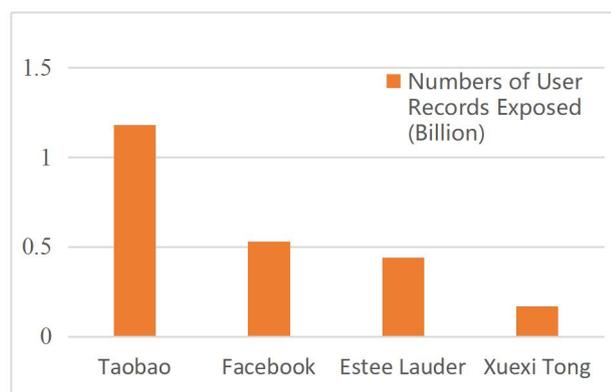

Figure 1

Several typical user data exposures in 2020-2021.

In fact, in the Internet age, many well-known companies have leaked their customers' privacy. Xuexi Tong, a study software for Chinese college students, was reported to have forced students to open some permissions and sell more than 170 million user data (Luo, 2022), and Estee Lauder's official website was hacked, which caused more than 440 million user data leaks (Arghire, 2020), etc.

These events sounded the alarm. Firstly, we would like to discuss the privacy leaks caused by the application operators, or app operators. On the one hand, in order to provide service with higher quality, app operators need to collect some personal information, including but not limited to, basic personal information, such as name, phone number, age, gender, region; and expanded personal information, such as political stance, religion, relationships, and so on. On the other hand, if this personal information is illegally sold by the app operators, or used for political purposes, it is undoubtedly a breach of personal privacy. In the age of big data, such leaks have become even more dangerous, because personal information is like an invisible and untouchable network, which can be explored from one point to another.

Secondly, there are situations from outside the app operators, such as hacking. Due to insufficient database protection measures, many enterprises have encountered attacks from hackers, which brings huge losses to enterprises, and also threatens the security of user privacy. On the one hand, user privacy information has become a convenient tool for earning money for regular service industry and subterranean economical industry; on the other hand, the Internet makes user privacy spread more rapidly, making the scope and impact of privacy data leakage increasing. Due to technical reasons, many software on the market still has some security vulnerabilities. These vulnerabilities may allow hackers to access personal information in the database.

Some mobile apps forcibly require a lot of permissions unrelated to the service it provided, such as access to mobile photo albums, access to address books, access to cameras, access to recording devices, etc. In order to explore this phenomenon, we installed several apps said to be with mandatory permissions, and found that some software directly quit unexpectedly after users refuse to provide those permissions, so that users cannot continue to use the apps. Some are more subtle, such as adding some clauses to the Privacy Agreement, which the average users will not check carefully. This allows some mobile apps to collect and sell users' private information.

Users' private information may flow to subterranean markets and be sold to intermediaries, who resell them to other companies. These companies, including real estate and insurance companies, buy users' privacy information, such as names, phone numbers, addresses, and even ID numbers, family conditions, financial status, and business fields, then regard those users as potential customers. This can lead to users receiving many "harassing phone calls." What is more dangerous is that the personal information may also flow to fraud companies, who make "fraud phone calls" to users, posing the risk of financial losses to them.

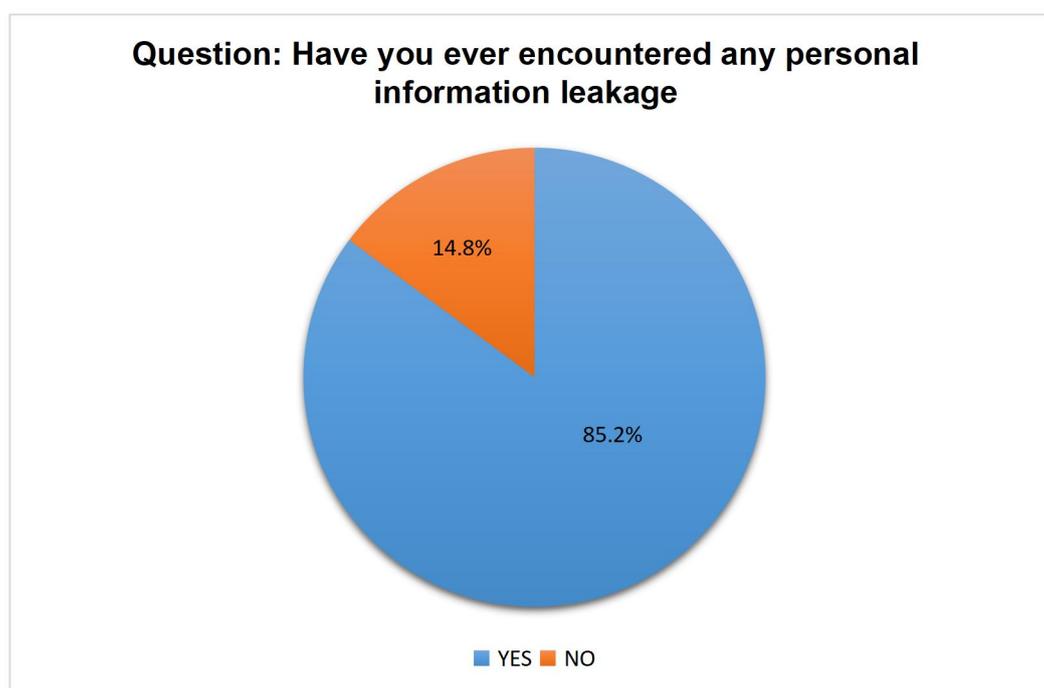

Figure 2　*Source*: China Consumers' Association.

**Identity Documents and Personal Privacy**

A Passport is a common personal identification document. In 1998,

electronic chip technology was introduced into the passport making process by the ICAO. Currently, many countries use e-passports, which contain a chip that records a lot of personal privacy information, even photo and fingerprint information. This is to prevent counterfeiting and improve the efficiency and convenience of passports, but it also leads to potential risks of privacy leakage. Under certain conditions, hackers can copy the information from the electronic chips, and the biological information is also at risk of being falsified (Li, 2013). Information in the chips of the passports in some countries or regions can be read directly even the passports are closed, while that of the passports in some other countries or regions can not be read the when closed, because those passports have a design which can block the external side recording or stolen reading, effectively protecting personal privacy. For instance, Australian passports have a screen chip page design that needs to open to a specific page number to remove the signal shielding, using the machine reading function.

We hope that more countries and regions will pay attention to the risk of privacy leakage in passports, and develop corresponding technologies to reduce the risk. The Extended Access Control (EAC) of the second-generation e-passport has a big security hazard, and the Supplemental Access Control (SAC) that is about to be applied to the third-generation e-passport will help alleviate this problem (Liu et al., 2015). The RFID technology used in the e-passport itself has certain risks, so the chip should be encrypted in multiple layers to prevent illegal reading and tampering.

Besides, bus cards, driving licenses, and other personal documents that are frequently used in daily life will also pose the risk of privacy leakage. Due to technology and cost constraints, these aspects of privacy protection is more dependent on the proper custody of personal documents by the holders. If conditions permit, privacy technology in passports can also be introduced into these documents. Or related companies can try to develop lower-cost privacy protection measures to protect users' privacy. In addition, relying on technology development, some other technologies may come true in the future. For example, the chip in personal certificate can be automatically locked at more than a certain distance from the certificate holder; each time when the information in the chip is read requires the permission of the holder; and so on.

## Privacy Power Legislation

Throughout history, the right to privacy has long been an important part of human rights. According to *Universal Declaration of Human Rights* promulgated by the United Nations, "no one shall be subjected to arbitrary interference with his privacy, family, home or correspondence … Everyone has the right to the protection of the law against such interference or attacks."

In China, *the Civil Code of the People's Republic of China*, which began implementation in 2021, further strengthens privacy protection. "No organization or individual may infringe upon the privacy of others by means of stabbing, intrusion, disclosure or disclosure." The law strengthens the protection of Chinese citizens' privacy rights.

In the United States, the legislative branches and executive branches should not keep secret personal information records, and that individuals should have the right to know the use of their personal information recorded by those departments (Zhang, 2012).

In some countries, such as the United Kingdom, the right of privacy is not regarded as an independent personality right, but is attached to other personality rights, such as the right of portrait, reputation, etc. They believe that personal privacy is a thing outside the law or an ancillary value. Only after the existence of other personality rights are violated and other causes of action, they can undertake the infringement of privacy right prosecution. Because these countries have relatively weak protection of personal privacy, violations of privacy often occur.

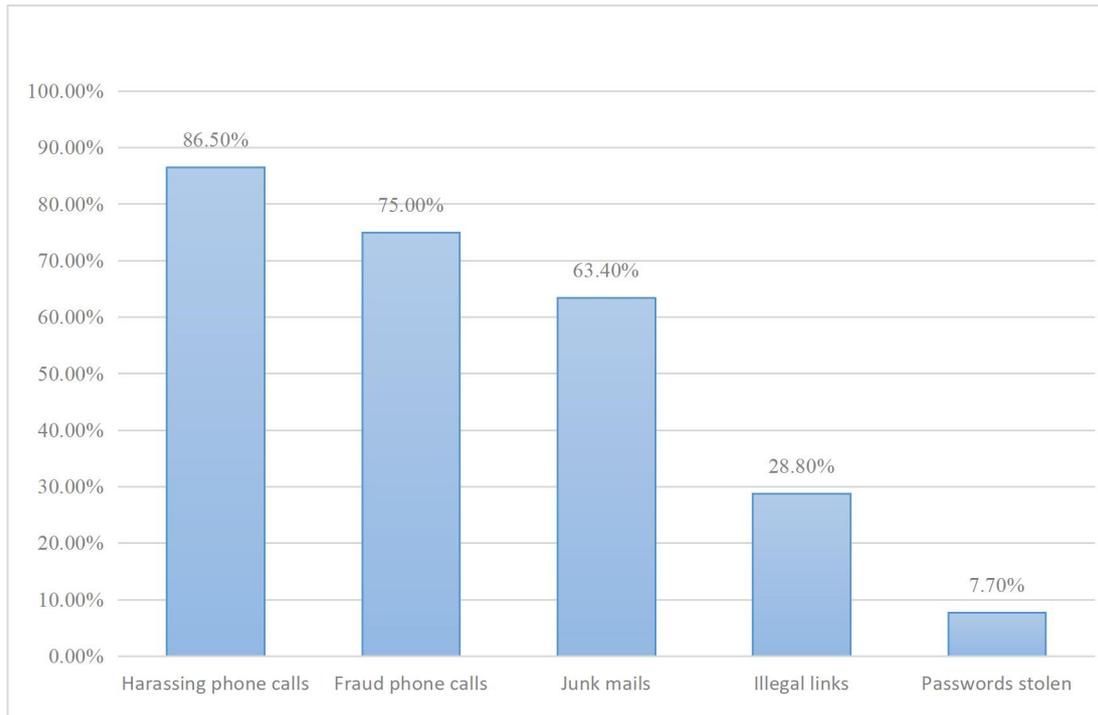

Figure 3    Personal information leakage behaviors. *Source*: China Consumers' Association.

**Possible Solutions to the Problem**

The relevant technology shall be updated. As we mentioned in the information confidentiality technology of the electronic passport, companies should try to develop and apply new technologies. Then, the Internet industry should establish feasible and uniform privacy protection standards. If necessary, this standard can require the participation of the legislative departments to make it more unified and widely used.

In addition, enterprises should also strengthen the database protection measures and data encryption measures. This can reduce the risk of privacy breach resulted from hacking. Enterprises should not ask for unnecessary private information and permissions from users.

The government should improve relevant laws and regulations through legislative form, strengthening the protection of citizens' privacy rights. It is suggested to further clarify the rights and obligations of consumers and service providers in Internet services, especially restrictions on the obligations and responsibilities of app service providers. Moreover, evaluating the response measures and risks of privacy leakage properly in personal information and data applications will count as well.

Citizens themselves should raise their awareness of privacy protection. Firstly, citizens should access their services from formal channels. Secondly, citizens should not arbitrarily give the app unnecessary permission and should not enter private information at will. Finally, when their privacy information is leaked, consumers should use the appropriate means to protect their legitimate rights and interests.

**CONCLUSION**

In today's digital age, the problem of privacy leakage is growing. The mobile phone apps that we use, the website accounts that we registered, and the identity documents that we use, can all be the culprits in leaking our private information. But at the same time, the relevant technology is also improving, which is having great implications for privacy protection. In addition, Internet companies, government departments, and individuals should all take measures, making the extensive participation and joint governance of all sectors of society protect our private information.

**LIMITATIONS**

Several common aspects of the existence of privacy risks are discussed in this essay, but some limitations remain. First of all, there are many reasons and ways for privacy leakage. In addition to the several aspects mentioned in this article, there are also many ways, such as the video surveillance systems in hotels being invaded. It is very difficult to discuss these pathways one by one. Second, considering the practical reasons, the practical solution to the privacy leakage problem is more complex than the theory mentioned in the article.

**REFERENCES**

[1] Arghire, I. (2020). Beauty and the Breach: Estée Lauder Exposes 440 Million Records in Unprotected Database, *Security Week*, 11 February 2020. Retrieved from https://www.securityweek.com/beauty-and-breach-est%C3%A9e-lauder-exposes-440-million-records-unprotected-database/

[2] China Consumers' Association, *App Personal Information Leakage Investigation Report*, 29 August 2018. Retrieved from https://www.cca.org.cn/jmxf/detail/28180.html?tdsourcetag=s_pctim_aiomsg/

[3] Li, W. (2013). Some Thoughts Based on the Issuance of the New Electronic Passport in China. *Science and Technology Innovation Guide*, 240-241.

[4] Liu, T., Zhang, D., & Jiang, Y. (2015). An E-Passport Security Review. *Computer Application and Software*.

[5] Louise, T. et al. (2022). A Framework for Data Privacy and Security Accountability in Data Breach Communications. *Computers and Security*, pp. 102657.

[6] Luo, Y. (2022). Xuexi Tong Student Information Leakage Incident Tracking: Some Seller Sold Overnight, Claiming to be Bought Out, *Beijing News*, 22 June 2022. Retrieved from https://finance.china.com/tech/13001906/20220622/37277003.html/

[7] United Nations, (1948). *Universal Declaration of Human Rights*, 10 December 1948.

[8] Wang, L. (1995). American Privacy Law and the Mass Media. *Journalism University* (01), 43-46.

[9] Zhang, F. (2012). The American Personal Information Protection System, *Beijing Court Web*, 03 May 2012. Retrieved from https://bjgy.bjcourt.gov.cn/article/detail/2012/05/id/885892.shtml/